\documentclass[useAMS,usenatbib,onecolumn,usegraphicx]{mn2e}

\newcommand{\lsim}{\, \, \raisebox{-0.8ex}{$\stackrel{\textstyle <}{\sim}$ }}

\newcommand{\beq}{\begin{equation}}
\newcommand{\eeq}{\end{equation}}
\newcommand{\beqar}{\begin{eqnarray}}
\newcommand{\eeqar}{\end{eqnarray}}

\title[Neutron star double-layer oscillations]  
{Astrophysical implications of double-layer torsional oscillations in a 
neutron star crust as a lasagna sandwich}
\author[H. Sotani, K. Iida, \& K. Oyamatsu]
{Hajime Sotani$^1$ \thanks{E-mail:sotani@yukawa.kyoto-u.ac.jp},
Kei Iida$^2$, and
Kazuhiro Oyamatsu$^3$
\\
$^1$Division of Science, National Astronomical Observatory of Japan, 2-21-1 Osawa, Mitaka, Tokyo 181-8588, Japan\\
$^2$Department of Mathematics and Physics, Kochi University, 2-5-1 Akebono-cho, Kochi 780-8520, Japan\\
$^3$Department of Human Informatics, Aichi Shukutoku University, 2-9 Katahira, Nagakute, Aichi 480-1197, Japan}

\begin{document}
\maketitle
\label{firstpage}

\begin{abstract}
In the crust of a neutron star, global torsional oscillations 
could occur in two elastic layers.  The outer and inner layers are 
composed of spherical and cylindrical nuclei and of cylindrical holes (tubes) 
and spherical holes (bubbles), respectively, while between these 
two layers, a phase of slab-like (lasagna) nuclei with vanishingly 
small elasticity is sandwiched.  In this work, we update systematic 
calculations of the eigenfrequencies of the fundamental oscillations in
the inner layer by newly allowing for the presence of tubes.  We find that 
the frequencies still depend strongly on the slope parameter of 
the nuclear symmetry energy, $L$, while being almost independent of the 
incompressibility of symmetric nuclear matter.  We also find that the 
fundamental frequencies in the inner layer can become smaller than 
those in the outer layer because the tube phase has a relatively 
small shear modulus and at the same time dominates the inner layer in 
thickness. As a result, we can successfully explain not only the 
quasi-periodic oscillations originally discovered in the observed X-ray 
afterglow of the giant flare of SGR 1806--20 but also many others 
recently found by a Bayesian procedure. 
\end{abstract}


\section{Introduction}
\label{sec:I}

Neutron stars, which are stellar remnants of core-collapse supernova 
explosions that occur at the last moment of massive stars, are 
composed of matter under extreme conditions, namely, such high 
density and large neutron excess as to be very difficult to realize on earth. 
In fact, the density inside the star can significantly exceed the 
normal nuclear density under strong gravitational field,  
while the neutron excess can become extremely large under charge neutral 
and beta equilibrium conditions \citep{NS}.  Thus, observations of
neutron stars are expected to help us to probe the physics under such 
extreme conditions, particularly given the difficulty in terrestrial 
laboratories in obtaining relevant information about matter in
neutron stars.  Even at the present time when information from two solar
mass neutron stars and a binary neutron star merger is available \citep{A2018},
however, the equation of state (EOS) of neutron star matter and 
hence the neutron star structure are still uncertain.

In spite of the uncertainties in the EOS of neutron star matter, 
a schematic view of neutron star structure can be drawn 
as follows.  Under the envelop composed mostly of a liquid 
metal, the matter is considered to have a lattice structure via the 
inter-ionic Coulomb interaction.  This crystalline region is 
referred to as a crust.  In the deepest region of the crust, 
below which the matter becomes uniform and constitutes a core, nuclei present
are so closely packed that the nuclear shape, which is normally roughly
spherical, could turn to cylinder (``spaghetti''), slab (``lasagna''), tube or 
cylindrical hole (``ani-spaghetti''), and bubble or spherical hole (``Swiss 
cheese'') as the density increases.  
Such exotic shapes are often called nuclear pasta \citep{LRP1993,O1993}.
This nuclear pasta is embedded in a gas of dripped neutrons and thus
can be viewed as a liquid-gas mixed phase of nuclear matter.  Since the
crystalline order of the phases of cylindrical nuclei, slab-like nuclei, 
and tubes is low-dimensional, furthermore, these phases are liquid 
crystals \citep{PP1998}.  Interestingly, it is known that the appearance 
of pasta structures depends strongly on the slope parameter of nuclear 
symmetry energy \citep{OI2007}, of which the determination is one of
the important problems in nuclear physics \citep{L2017}.
Observational evidence for the presence of nuclear pasta would thus be 
highly desired.

To extract information of neutron star interiors from observations,  
asteroseismology is a very powerful technique, just like the 
seismology for the Earth and the helioseismology for the Sun. 
That is, since the characteristic frequencies observed from 
neutron stars may well be more or less related to the interior 
properties, one could obtain the interior information by somehow
observing such frequencies, identifying them as eigenfrequencies of
some global oscillation modes, and then solving an inverse problem.  
Such frequencies could be obtained from gravitational waves that 
would radiate from the interiors and reach us due to the strong permeability.  
In fact, many possibilities of extracting the 
neutron star properties via direct detection of the gravitational waves
have been proposed (e.g., \cite{AK1996,STM2001,SKH2004,SYMT2011,DGKK2013}). 
Study in this direction is so promising as to make us expect to obtain
important information on the neutron star interiors in the near future.

As long as neutron star asteroseismology is concerned,
quasi-periodic oscillations (QPOs) in X-rays have been only known
electromagnetic signals of global oscillations.  Up to now, three 
giant flares have been observed from different soft gamma repeaters (SGRs).
In two of them, namely, SGR 1900+14 and SGR 1806-20, several QPOs have 
been found in the X-ray afterglow following the respective
giant flare, where the observed QPO frequencies are in the range of tens of Hz 
up to kHz \citep{I2005,SW2005,SW2006}.  In SGR 1806-20, another QPO, 
i.e., the 57 Hz QPO, was also found from the shorter and less energetic 
recurrent 30 bursts \citep{QPO2}.  Since the central object in the SGR is 
considered to be a strongly magnetized neutron star, the observed QPOs 
may well come from the global oscillations of the 
neutron star.  Given that typically, the frequencies induced by 
acoustic oscillations in the star are around kHz \citep{VH1995}, one
has difficulty in identifying the QPOs of frequency lower than 
$\sim 100$ Hz as the acoustic oscillations.  In practice, it is
generally accepted that the mechanisms for explaining such lower QPO 
frequencies are either the crustal torsional oscillations, the 
magnetic oscillations, or the coupled modes (magneto-elastic oscillations).

However, calculations of the magnetic oscillation frequencies suffer
several uncertainties.  The geometry and strength distribution of the 
internal magnetic fields are poorly known, although the magnetic 
oscillations depend strongly on such magnetic structure 
\citep{GCFMS2013}.  In addition, one has to take into account the 
uncertain core EOS if the magnetic fields penetrate into 
the core region.  To avoid such uncertainties, in this study we focus on the 
crustal torsional oscillations by ignoring the coupling to the magnetic
oscillations.  Note that even in the absence of this coupling, the calculated 
eigenfrequencies of the crustal torsional oscillations are still controlled by 
several physical parameters that are relatively well-known but not yet
determined, i.e., such crustal properties as the shear modulus and the 
superfluid density, as well as the star's radius $R$ and mass $M$.  By 
identifying the observed QPO frequencies as such eigenfrequencies, 
therefore, one can manage to constrain the crustal properties \citep{SA2007,SW2009,GNJL2011,SNIO2012,PA2012,SNIO2013a,SNIO2013b,S2014,S2016,SIO2016,SIO2017a,SIO2018}.

In most of these earlier studies of the crustal torsional 
oscillations, it was assumed that only the phase of spherical nuclei 
oscillates quasiperiodically, while the pasta phases remain free from 
oscillations.  Since most of the pasta phases are liquid crystals, however, 
their elastic properties could be responsible for global oscillations.  In  
contrast to a naive view that the shear modulus decreases continuously 
in the pasta phases and eventually vanishes at the crust-core boundary,
which was adopted in \cite{S2011,PP2016}, we have recently attempted
to introduce a more realistic effect of the pasta elasticity into 
the torsional oscillations \citep{SIO2017a,SIO2018}.  In this attempt,  
it was noted that for slab-like nuclei, the transverse shear response 
vanishes for long-wavelength perturbations \citep{dGP1993,PP1998}. 
That is, within the linear analysis, the phase of slab-like nuclei 
behaves as a fluid. This indicates that the torsional oscillations 
that could be excited within the phases of spherical and cylindrical 
nuclei would be separable from those within the phases of 
tubes and bubbles.

In our recent study \citep{SIO2018}, we calculated eigenfrequencies
of the torsional oscillations that occur inside the phases of 
spherical and cylindrical nuclei and showed that the QPO frequencies observed 
in SGR 1806-20 and SGR 1900+14, except for the 26 Hz QPO observed in SGR 
1806-20, can be explained in terms of such oscillations.  Additionally, 
since the torsional oscillations are supposed to be confined 
within a layer composed of spherical and cylindrical nuclei, we 
discussed the overtone torsional oscillations, which have radial 
nodes in such a manner that is dependent on the thickness of the layer.  By 
identifying the kHz QPO observed in SGR 1806-20 as the 1st overtone 
torsional oscillation, we attempted to constrain the incompressibility 
of symmetric nuclear matter for given $M$ and $R$.  By combining the 
resultant constraint with the constraint from empirical data for nuclear 
giant monopole resonances, furthermore, not only did we manage to 
constrain $M$ and $R$, but we obtained a more severe constraint on the 
slope parameter $L$ of nuclear symmetry energy.

Even before our previous work \citep{SIO2018}, we suggested the 
possibility that the 26 Hz QPO in SGR 1806-20 stems from torsional 
oscillations that occur only in a deeper layer of the crust than the
slab phase, i.e., in a layer of composed of tubes and bubbles.  As a first 
step \citep{SIO2017a}, we focused on the torsional oscillations in the 
bubble phase alone by simply assuming zero elasticity in the tube phase.
It was noted that the lowest fundamental frequency in the bubble 
phase could explain the 26 Hz QPO because the enthalpy density is 
relatively small in the bubble phase.  In this work, by taking into 
account the effect of the tube phase, we will give a more realistic 
evaluation of the eigenfrequencies of torsional oscillations that occur
in the tube-bubble layer and thereby examine whether one could still 
explain the 26 Hz QPO.  Within the same framework, moreover,
we will discuss possible identification of newly found QPOs in SGR 
1806-20 by a Bayesian procedure \citep{MCS18}.

In Sec.\ \ref{sec:II}, we summarize a model for the neutron star crust 
that is constructed in such a way as to depend on the EOS of 
nuclear matter.  Section \ref{sec:III} is devoted to description of the 
shear modulus that is consistent with the crust model summarized in Sec.\ 
\ref{sec:II}.  In Sec.\ \ref{sec:IV}, we calculate the eigenfrequencies of 
fundamental shear torsional oscillations in two elastic layers within the 
crust and compare them with the low-lying QPO frequencies observed from 
SGR 1806--20.  Finally, concluding remarks and details of such comparison 
are given in Sec.\ \ref{sec:V} and Appendix \ref{sec:appendix_1}, respectively. 
Throughout the text, we use units in which $c=G=1$, where $c$ and $G$ denote 
the speed of light and the gravitational constant, respectively.

\section{Model for neutron star crust}
\label{sec:II}

We start with construction of a neutron star crust in a spherically
symmetric configuration. This is because for neutron stars observed as SGRs
the magnetic and rotational energies are much smaller than the 
gravitational binding energy \citep{K1998,H1999}.  Then, the crust can be
constructed by integrating the Tolman-Oppenheimer-Volkoff (TOV) equation 
together with the EOS of matter in the crust. Correspondingly, the 
metric is given in spherical polar coordinates as
\begin{equation}
  ds^2 = -e^{2\Phi(r)} dt^2 + e^{2\Lambda(r)} dr^2 + r^2 d \theta^2 + r^2\sin^2\theta d\phi^2,
\end{equation}
where $\Lambda(r)$ is directly connected to the mass function, $m(r)$, via 
$\exp(-2\Lambda)=1-2m/r$.

It is advantageous that we dispense with the core EOS, which is 
significantly uncertain.  In integrating the TOV equation, therefore,
we set the values of $R$ and $M$ and then go inward from the star's 
surface down to the crust-core boundary \citep{IS1997}.

To construct the crust in equilibrium, one has to prepare the EOS of
crustal matter that is in beta equilibrium and globally charge neutral.
This EOS can in turn be constructed in such a way that is dependent on 
the bulk energy of zero temperature nuclear matter per baryon, which can 
generally be expanded in the vicinity of the saturation point of
symmetric nuclear matter with respect to the baryon number density 
($n_{\rm b}$) and the neutron excess ($\alpha$) (see \cite{L1981}):
\begin{equation}
  w(n_{\rm b}, \alpha) = w_0  + \frac{K_0}{18n_0^2}(n_{\rm b} - n_0)^2 + \left[S_0 + \frac{L}{3n_0}(n_{\rm b} - n_0)\right]\alpha^2. \label{eq:w}
\end{equation}
Here $w_0$ and $K_0$ are the bulk energy and the incompressibility of 
the symmetric nuclear matter at the saturation density of $n_{\rm b}=n_0$, 
while $S_0$ and $L$ are the parameters that characterize
the nuclear symmetry energy, $S(n_{\rm b})$, i.e., $S_0=S(n_0)$ and 
$L=3n_0(dS/dn_{\rm b})$ at $n_{\rm b}=n_0$.  Among these five saturation
parameters, $n_0$, $w_0$, and $S_0$ are fairly well
constrained from empirical data for masses and charge radii of stable
nuclei. On the other hand, the constraint on the remaining two parameters, 
$K_0$ and $L$, are relatively more difficult to obtain, because 
these are related to the density change from $n_{\rm b}=n_0$.  In this 
study, therefore, we adopt the phenomenological EOSs of crustal
matter that were constructed by \cite{OI2003,OI2007} in such as way as to 
depend on $K_0$ and $L$ (hereafter refereed to as OI-EOSs). These
EOSs allow us to systematically examine the dependence of the crustal 
oscillations on $K_0$ and $L$.

Let us briefly summarize the OI-EOSs. The expression for the 
energy of bulk nuclear matter used in the OI-EOSs was
constructed in a Pade form with respect to the density and in a 
parabolic approximation with respect to the neutron excess, and fitted to
empirical data for masses and charge radii of stable nuclei within the 
Thomas-Fermi approach \citep{OI2003}.  Consequently, the saturation 
parameters in Eq.\ (\ref{eq:w}) were given for more than 200 sets
of $K_0$ and $y\equiv -K_0S_0/(3n_0L)$.  Then, within the Wigner-Seitz 
approximation for five nuclear shapes, i.e., sphere, cylinder, slab, 
tube, and bubble, the equilibrium nuclear shape and size in the crust 
were determined as a function of $n_{\rm b}$ by optimizing the  
energy density functional in the presence of a neutralizing
uniform electron gas and a gas of dripped neutrons~\citep{OI2007}. 
In this study we confine ourselves to several sets of the saturation
parameters, which cover not only typical but also extreme cases as in
Table~\ref{tab:EOS}.  We remark that the typical values are 
empirically deduced as, e.g., $30\lsim L \lsim 80$ MeV \citep{Newton2014} 
and $K_0=230\pm 40$ MeV~\citep{KM2013} or $250 \lsim K_0 \lsim 315$ 
MeV~\citep{SSM2014}.

Since we focus on the torsional oscillations that are trapped inside 
the phases of tubes and bubbles in this study, we also show the 
transition densities from the slab to the tube phase (S-CH), 
from the tube to the bubble phase (CH-SH), and from the bubble to 
the uniform phase (SH-U) in Table~\ref{tab:EOS}.   As already 
predicted by \cite{OI2007}, the pasta structure is more
difficult to appear for larger $L$.  In fact, some of the
pasta structures are predicted to disappear for the cases with 
$L=76.4$ and 146.1 MeV, which are denoted by the asterisk 
in the column of $K_0$ in Table~\ref{tab:EOS}.  We remark that 
the thickness of each pasta phase strongly depends on not only $K_0$ 
and $L$ but also the stellar compactness ($M/R$) \citep{SIO2017b}.
We also remark that the transition densities tabulated in 
Table~\ref{tab:EOS} are not obtained at constant pressure; in a 
real situation, the density jumps at the transition pressures, but this
jump is tiny because the transitions are of weak first order.

\begin{table}
\centering
\begin{minipage}{100mm}
\caption{
The transition densities at the S-CH, CH-SH, and SH-U boundaries are shown for 
several sets of the OI-EOSs characterized by $K_0$ and $L$.  In the 
cases in which the asterisk is affixed to the value of $K_0$, some 
of the pasta phases are not predicted to appear.  The values with $*1$ 
and $*2$ denote the transition densities from the cylindrical-hole to the 
uniform phase and from the phase with spherical nuclei to the 
uniform phase, respectively.
}
\begin{tabular}{cc|cccc}
\hline\hline
  $K_0$ (MeV) & $L$ (MeV) & S-CH (fm$^{-3}$) & CH-SH (fm$^{-3}$) & SH-U (fm$^{-3}$)  \\
\hline
  180 & 17.5 & 0.09811 & 0.10206  &  0.10321     \\  
  180 & 31.0 & 0.08739 & 0.09000   & 0.09068     \\  
  180 & 52.2 & 0.07733 & 0.07885   & 0.07899     \\  
  230 & 23.7 & 0.09515 & 0.09817  &  0.09866     \\  
  230 & 42.6 & 0.08411 & 0.08604   & 0.08637     \\  
  230 & 73.4 & 0.07284 & 0.07344   & 0.07345     \\  
  360 & 40.9 & 0.09197 & 0.09379  &  0.09414     \\  
  $^*$360 & 76.4   & 0.07890 & ---   & 0.07918$^{*1}$     \\  
  $^*$360 & 146.1 & --- & ---   & 0.06680$^{*2}$     \\  
\hline\hline
\end{tabular}
\label{tab:EOS}
\end{minipage}
\end{table}

In considering the torsional oscillations, furthermore, the 
effective enthalpy, $\tilde{H}$, that participates in the oscillations is
another important factor, because the shear velocity $v_s$ is given by 
$v_s^2=\mu/\tilde{H}$, where $\mu$ is the shear modulus to be discussed 
in the next section, and because the fundamental frequency of the 
torsional oscillations is proportional to $v_s$ \citep{HC1980}.  In practice, 
for the torsional oscillations in the tube and bubble phases, the effective
enthalpy can be expressed as
\begin{equation}
  \tilde{H} = \frac{N_i + {\cal R}(A - N_i)}{A}H,  \label{eq:H}
\end{equation}
where $N_i$ denotes the number of neutrons inside a single tube or 
bubble, $A$ is the total nucleon number in a Wigner-Seitz cell, and $H$ 
denotes the local enthalpy given by $H=\varepsilon + p$ with the energy density 
($\varepsilon$) and pressure ($p$).  The coefficient ${\cal R}$ is a parameter 
that characterizes a participant ratio, i.e., how much ratio of nucleons
outside the tube or bubble comove with it non-dissipatively via 
entrainment, namely, Bragg scattering off the corresponding lattice.  Note 
that the non-participant nucleons behave as a superfluid.  There are two 
extreme cases: All the nucleons inside the Wigner-Seitz cell contribute 
to the effective enthalpy for ${\cal R}=1$ (maximum enthalpy), while 
no nucleons outside the tube or bubble do so for ${\cal R}=0$ (minimum 
enthalpy).  In general, ${\cal R}$ has an intermediate value that depends 
on the band structure and pairing gap for the nucleons outside the tube or 
bubble and hence changes with density.  In this study, we simply 
consider only the extreme cases in which ${\cal R}$ is constant at 
1 or 0 in the whole region of the tube and bubble phases.  We remark that the 
value of ${\cal R}$ in the bubble phase is predicted to be 
$\sim 0.34-0.38$ at $n_{\rm b}=0.08$ fm$^{-3}$, according to the band 
calculations by \cite{Chamel2012}.  Incidentally, we naively assume that 
all the $N_i$ neutrons comove with the interface of the tube or bubble, just 
like bubbles in boiled water.  This might not be always the case with the 
superfluid system considered here in which a non-dissipative hydrodynamic flow 
could arise in such a way that some neutrons go across the interface 
\citep{MU2016}.

\section{Shear modulus}
\label{sec:III}

Let us now proceed to the shear modulus, $\mu$, which is 
associated with the distortion energy to be produced
by adding long-wavelength transverse shear deformation 
of each of the five phases of inhomogeneous nuclear matter. 
The distortion energy comes mainly from the change of the Coulomb 
energy due to the deformation, and a particular form of
the corresponding shear modulus was adopted in our previous
studies except for the tube phase.
In the case of a bcc Coulomb lattice composed of 
spherical nuclei, the effective shear modulus was 
originally derived as 
\begin{equation}
  \mu_{\rm sp} = 0.1194\frac{n_i(Ze)^2}{a},     \label{eq:musp}
\end{equation}
where $n_i$, $Z$, and $a$ denote the ion number density, 
the charge number of the nuclei, and the radius of the 
Wigner-Seitz cell, i.e., $n_i=(4\pi a^3/3)^{-1}$ 
\citep{OI1990,SHOII1991}.  Note that this $\mu_{\rm sp}$ 
was obtained via Monte Carlo method by averaging over all 
directions of the wave vector of the distortion 
with the assumption that each nucleus is a point particle.  
Recently, this shear modulus has been
modified by taking into account the effect of electron 
screening \citep{KP2013} and the effect of polycrystalline
nature \citep{KP2015}. In this study, however, 
we adopt the traditional formula given by Eq.\ (\ref{eq:musp}) 
for simplicity.

The elastic properties in the rod and slab phases 
have been discussed by \cite{dGP1993,PP1998}.  The shear 
modulus in the phase of cylindrical nuclei was 
derived through the deformation energy to be produced 
by adding a two-dimensional displacement perpendicular 
to the elongated direction of the equilibrium 
configuration of cylindrical nuclei.  In practice, it can 
be effectively expressed as
\begin{equation}
  \mu_{\rm cy} = \frac{2}{3}E_{\rm Coul} \times 10^{2.1(w_2-0.3)},   \label{eq:mucy}
\end{equation} 
where $E_{\rm Coul}$ and $w_2$ denote the Coulomb energy 
per volume of a Wigner-Seitz cell and the volume 
fraction of cylindrical nuclei, respectively, and 
the coefficient of $2/3$ comes from the average over 
all directions between the wave-vector of the distortion
and the elongated direction under the assumption that 
crystallites of cylindrical nuclei randomly point. 
We remark that 
in the liquid drop model $E_{\rm Coul}$ is given by 
\begin{equation}
  E_{\rm Coul} = \frac{\pi}{2} (\rho_p R_p)^2 w_2\left[\ln\left(\frac{1}{w_2}\right)-1+w_2\right],
  \label{eq:E_coul}
\end{equation}
where $\rho_p$ and $R_p$ are the proton charge density and the proton radius
of a cylindrical liquid drop \citep{RPW1983}. 
By following a similar line of argument, 
it was shown that the deformation energy in the phase of 
slab-like nuclei becomes of higher order with respect 
to the displacement.  That is, this phase behaves as a 
fluid within the linear response.  This is the reason 
why one can consider the torsional oscillations inside the 
phases of spherical and cylindrical nuclei separately 
from those inside the phases of tubes and bubbles.

The shear modulus in the tube (bubble) phase, i.e., $\mu_{\rm ch}$ 
($\mu_{\rm sh}$), can be derived in a similar fashion to that 
in the phase of cylindrical (spherical) nuclei, because the liquid 
crystalline structure of tubes (bubbles) is the same as that in the phase
of cylindrical (spherical) nuclei.  In this study, therefore, we adopt 
Eq.\ (\ref{eq:mucy}) for the tube phase and Eq.\ (\ref{eq:musp}) for the bubble
phase by properly replacing the relevant quantities in these 
formulae:  In the tube phase, $w_2$ in Eq.\ (\ref{eq:mucy})
(including $E_{\rm Coul}$) is replaced
by the volume fraction of a gas of dripped neutrons, while in the 
bubble phase $n_i$ and $Z$ are replaced by the number density of 
bubbles and the effective charge number $Z_{\rm bubble}$ of a bubble,
respectively \citep{SIO2017a}.  In practice, $Z_{\rm bubble}$ is given 
by $Z_{\rm bubble}=n_QV_{\rm bubble}$, with the volume of the bubble, 
$V_{\rm bubble}$, and the effective charge number density of the bubble, 
$n_Q$, defined by the difference of the charge number density 
inside the bubble from that outside the bubble, i.e., 
$n_Q=-n_{\rm e}-(n_{\rm p}-n_{\rm e})=-n_{\rm p}$ with the proton number density 
outside the bubble ($n_{\rm p}$) and the number density of a uniform electron 
gas ($n_{\rm e}$).

\begin{figure*}
\begin{center}
\begin{tabular}{cc}
\includegraphics[scale=0.5]{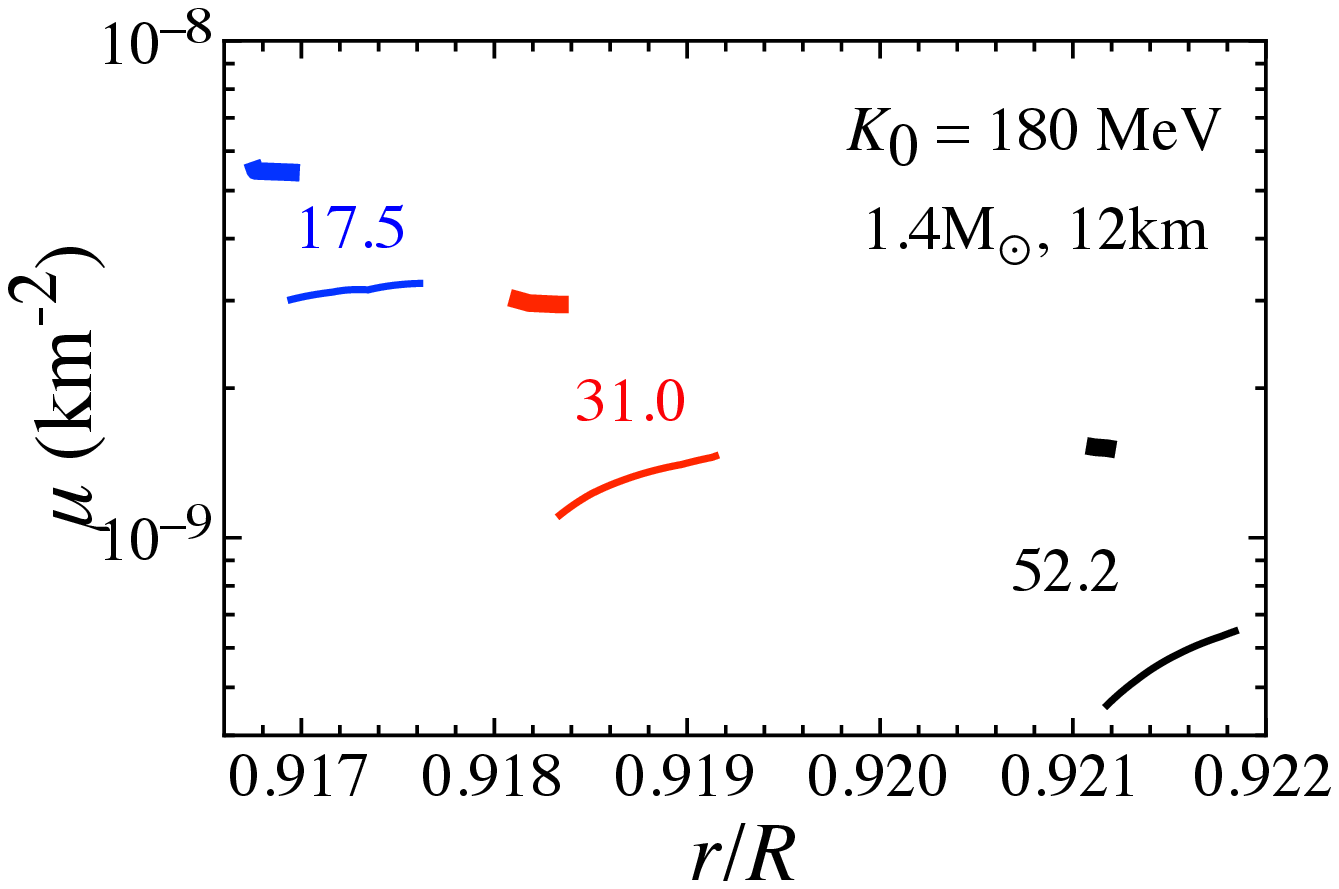} &
\includegraphics[scale=0.5]{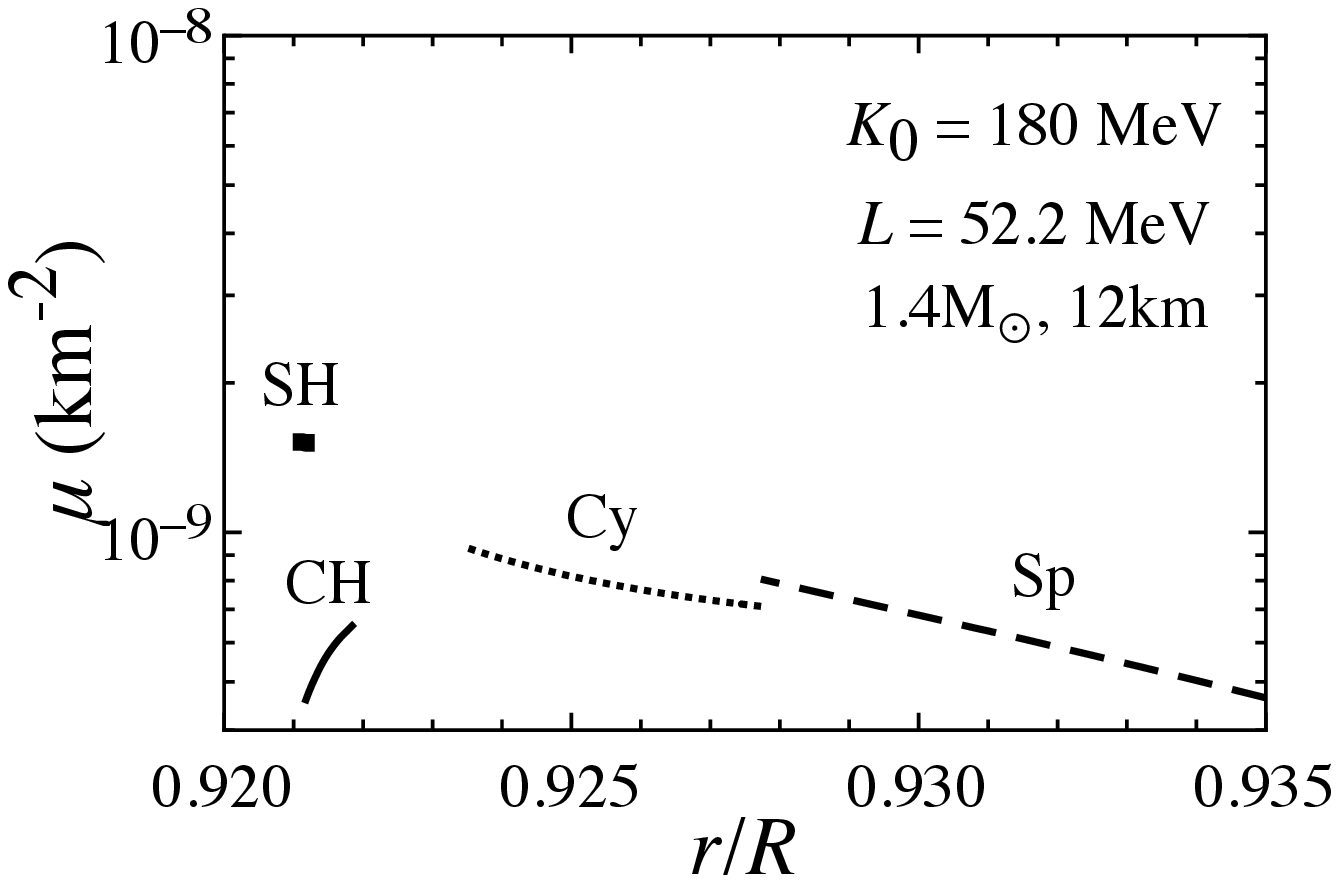} 
\end{tabular}
\end{center}
\caption{
(Color online) 
Left: Profile of the shear modulus in the tube phase (thin lines) and bubble phase 
(thick lines), calculated for the neutron star models with 
$M=1.4M_\odot$ and $R=12$ km.  Here, $K_0$ is fixed at 180 MeV,
while $L$ takes the value as labeled in the unit of MeV.
Right: For the neutron star model with $M=1.4M_\odot$ and 
$R=12$ km constructed with $K_0=180$ MeV and $L=55.2$ MeV, 
the profile of the shear modulus in
the phase of spherical nuclei (Sp) and in the phase of cylindrical nuclei (Cy)
is shown as well as that in the tube (CH) phase and the bubble (SH) phase.
}
\label{fig:mu}
\end{figure*}

In Fig.~\ref{fig:mu},  we illustrate the profile of the shear modulus inside the tube 
and bubble phases for neutron star models constructed from the first
three sets of the OI-EOSs listed in Table~\ref{tab:EOS}.   From this figure, 
one can observe that the shear modulus becomes discontinuous at the transition 
between the tube and bubble phases, which is similar to the case of the 
transition between the phases of spherical and cylindrical nuclei 
\citep{SIO2018}.   In addition, it is to be noted that the shear modulus 
in the tube phase can decrease as the density increases and that this
tendency becomes stronger for larger $L$.  This tendency may well 
come from the decrease of the volume fraction of a gas of dripped 
neutrons with density (e.g., \cite{WI2003}).

\section{Torsional oscillation frequencies and comparison with QPOs}
\label{sec:IV}

We now turn to evaluations of the eigenfrequencies of fundamental 
torsional oscillations in the sphere-cylinder and tube-bubble layers of the 
crust of a neutron star.  To this end, we start with the perturbation 
equation in a spherical coordinate system, which is given by linearizing 
the relativistic equation of motion that determines the torsional 
oscillations \citep{ST1983,Sotani2007} as
\begin{equation}
  {\cal Y}'' + \left[\left(\frac{4}{r} + \Phi' - \Lambda'\right)+\frac{\mu'}{\mu}\right]{\cal Y}' 
      + \left[\frac{\tilde{H}}{\mu}\omega^2e^{-2\Phi} - \frac{(\ell+2)(\ell-1)}{r^2}\right]e^{2\Lambda}{\cal Y}=0,
      \label{eq:perturbation}
\end{equation}
where ${\cal Y}$ denotes the Lagrangian displacement in the $\phi$ direction, 
while $\tilde{H}$ is the effective enthalpy given by Eq.\ (\ref{eq:H}). 
With the appropriate boundary conditions, the problem to solve becomes an 
eigenvalue problem, where $\omega$ is the eigenvalue.  Then, the  
eigenfrequency of the torsional oscillations $f$ is given by $f=\omega/(2\pi)$.
As for the boundary conditions relevant to the torsional 
oscillations that are excited inside the tube and bubble phases, 
there are two boundaries, namely, the boundary between the bubble phase 
and uniform matter, which corresponds to the inner boundary, and the boundary 
between the slab and tube phases, which corresponds to the outer 
boundary.  In practice, one has to impose the zero-traction conditions 
at the inner and outer boundaries, i.e., ${\cal Y}'=0$.  In addition, one has 
to impose the junction condition at the boundary between the tube and bubble 
phases, where the traction should be continuous, i.e.,
\begin{equation}
  \mu_{\rm ch}{\cal Y}' = \mu_{\rm sh}{\cal Y}'.
\end{equation}

\begin{figure*}
\begin{center}
\begin{tabular}{cc}
\includegraphics[scale=0.5]{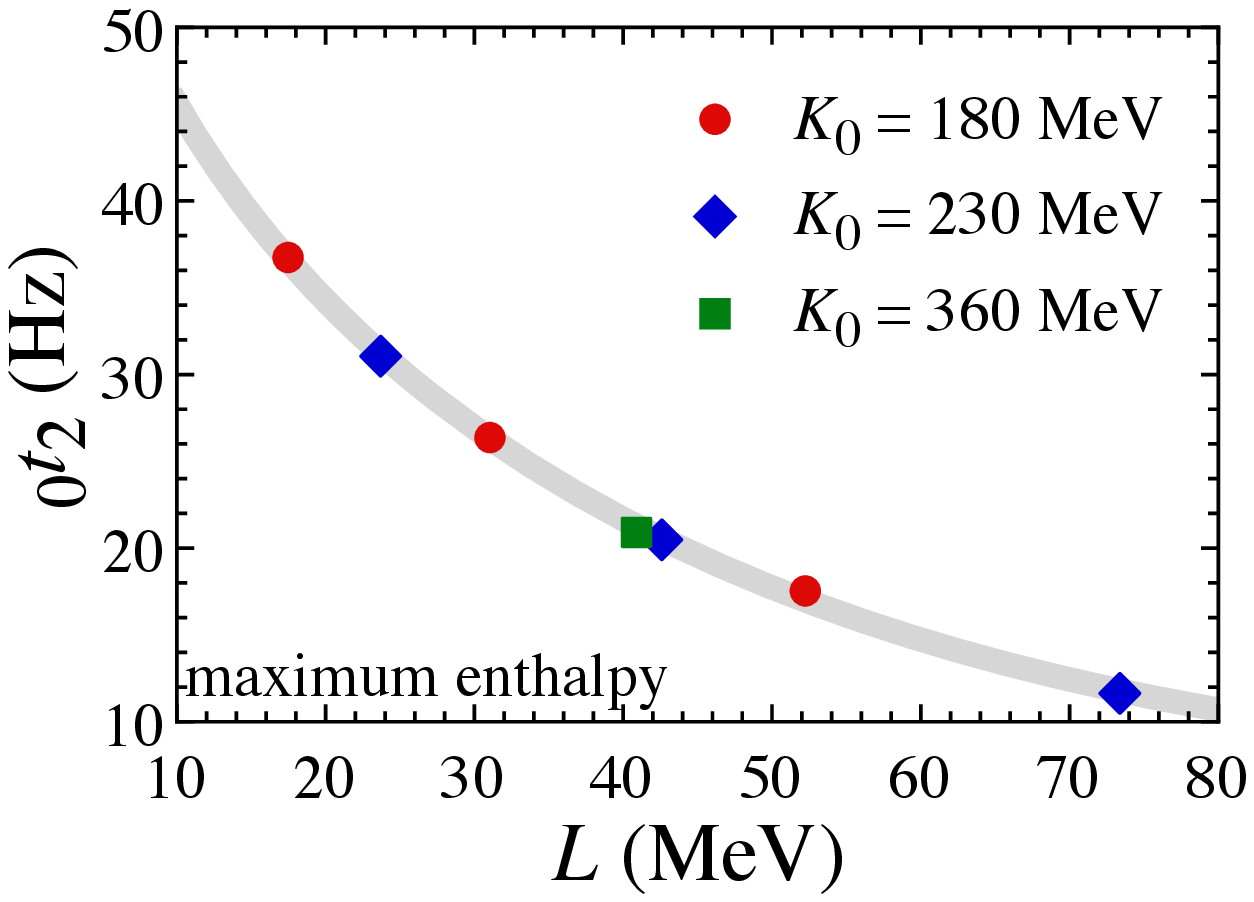} &
\includegraphics[scale=0.5]{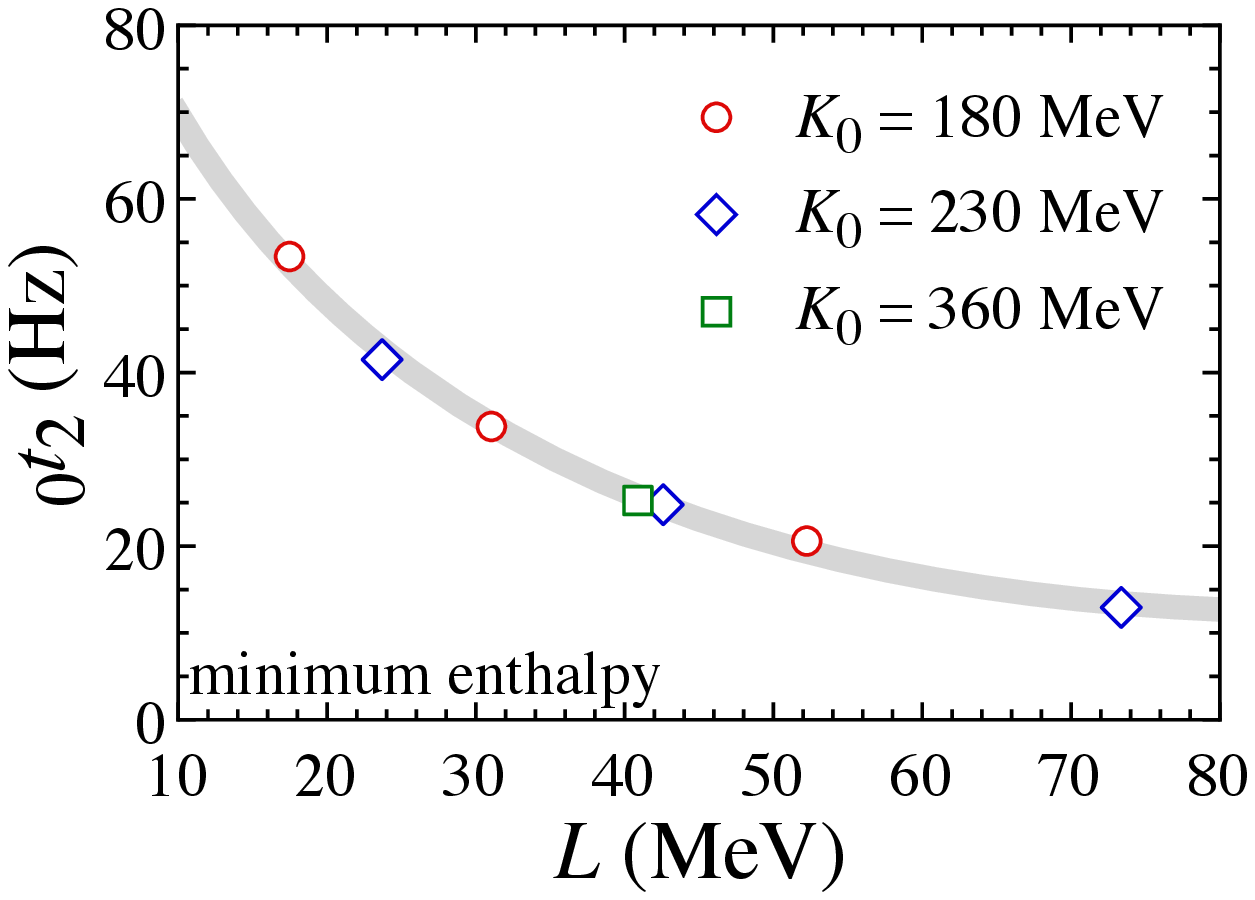}  
\end{tabular}
\end{center}
\caption{
(Color online)
Fundamental frequencies of the $\ell=2$ torsional oscillations in the tube 
and bubble phases, as obtained for the neutron star models with 
$M=1.4M_\odot$ and $R=12$ km as well as with various sets of $L$
and $K_0$.  Here, the left and right panels correspond to the results 
for the maximum and minimum enthalpies that participate in the 
oscillations, i.e., ${\cal R}=1$ and 0, respectively (see the text for 
details).  The solid line denotes the fitting given by Eq.\ (\ref{eq:fitting}).
}
\label{fig:t2L2-M14R12}
\end{figure*}

In Fig.~\ref{fig:t2L2-M14R12}, we plot the $\ell=2$ fundamental 
frequencies of torsional oscillations in the tube and bubble phases
that are calculated for the neutron star models with $M=1.4M_\odot$ and 
$R=12$ km, with various EOS parameter sets shown in 
Table~\ref{tab:EOS}, and with the maximum and minimum enthalpies
(${\cal R}=1$ and 0).  From this figure, one can observe that the frequency 
depends only weakly on $K_0$, but shows a significant sensitivity
to $L$.  In fact, we find that the $\ell=2$ fundamental frequencies can
be well fitted as a function of $L$ via
\begin{equation}
  {}_0t_2 = c_2^{(0)} + c_2^{(1)}\sqrt{L} + c_2^{(2)}L, \label{eq:fitting}
\end{equation}
where $c_2^{(0)}$, $c_2^{(1)}$, and $c_2^{(2)}$ are the adjustable 
parameters that depend on $M$ and $R$ as well as the value of 
${\cal R}$.  The obtained fitting [Eq.\ (\ref{eq:fitting})] is also 
shown in Fig.~\ref{fig:t2L2-M14R12}.   We remark that this fitting has
a different functional form from that obtained for the fundamental 
frequencies of crustal torsional oscillations in the phases of
spherical and cylindrical nuclei \citep{SIO2018}.  We also remark that the 
fundamental frequency in the tube and bubble phases can be smaller than 
that in the phases of spherical and cylindrical nuclei and that
the fundamental frequencies with general values of $\ell$ can 
also be well fitted in the same functional form:
\begin{equation}
  {}_0t_\ell = c_\ell^{(0)} + c_\ell^{(1)}\sqrt{L} + c_\ell^{(2)}L, 
\label{eq:fitting1}
\end{equation}
with a different set of the adjustable parameters $c_\ell^{(0)}$, 
$c_\ell^{(1)}$, and $c_\ell^{(2)}$.  Hereafter we will thus attempt to 
identify the observed QPO frequencies by using the fitting formula 
(\ref{eq:fitting1}) for the tube-bubble layer, in addition to 
the formula given in \cite{SIO2018} for the sphere-cylinder layer.

Note that the obtained fundamental frequencies in the tube-bubble
layer are generally lower than those obtained in our earlier 
analysis by assuming that only the bubble phase is oscillating 
\citep{SIO2017a}.  This tendency is more significant for larger values 
of $L$.  This is partly because for larger $L$, the bubble phase is 
less likely to appear, as shown in Table~\ref{tab:EOS}, and partly 
because the shear modulus and hence the shear velocity is relatively 
small in the tube phase, as shown in Fig.\ \ref{fig:mu}.  As we shall 
see later, therefore, the 26 Hz QPO observed from SGR 1806-20 is 
identified as the $\ell=4$ fundamental torsional oscillation in the 
tube-bubble layer, in contrast to our earlier analysis \citep{SIO2017a} 
in which it was identified as the $\ell=2$ fundamental torsional 
oscillation.

Up to now, we have already done many trials to identify the QPOs 
observed in SGR 1806-20 and SGR 1900+14 as the crustal torsional 
oscillations. As long as we adopt the QPO frequencies derived in the 
conventional non-Bayesian analyses of RXTE data for the X-ray afterglows 
of the giant flares and the recurrent X-ray outbursts, such identification 
has worked out relatively well
\citep{SNIO2012,SNIO2013a,SNIO2013b,SIO2016,SIO2017a,SIO2018}.
In fact, the observed QPOs, except for the 26 Hz QPO in SGR 1806-20, 
can be identified as the torsional oscillations inside the phases of 
spherical and cylindrical nuclei in such a way that the QPOs of 
frequencies lower than $\sim 200$ Hz correspond to the fundamental 
oscillations with various values of $\ell$, while the higher QPOs observed 
in SGR 1806-20 correspond to the overtones \citep{SIO2018}. 
In this case, since it is still uncertain how much fraction of dripped 
neutrons in the phase of cylindrical nuclei would be locked to the 
motion of protons in the nuclei, we introduced a parameter $N_s/N_d$, 
where $N_d$ and $N_s$ respectively denote the number of dripped neutrons 
outside the cylindrical nuclei and the number of a superfluid part 
of the dripped neutrons that behave independently of the oscillations, 
and examined the extreme cases with $N_s/N_d=0$ and 1.  We remark that 
all (none) of the dripped neutrons outside the cylindrical nuclei 
participate in the oscillations for $N_s/N_d=0$ $(1)$.  We also 
remark that for the corresponding value of $N_s/N_d$ in the phase of 
spherical nuclei, we adopt the results by \cite{Chamel2012}, which are 
based on the band theory.

\begin{figure}
\begin{center}
\includegraphics[scale=0.6]{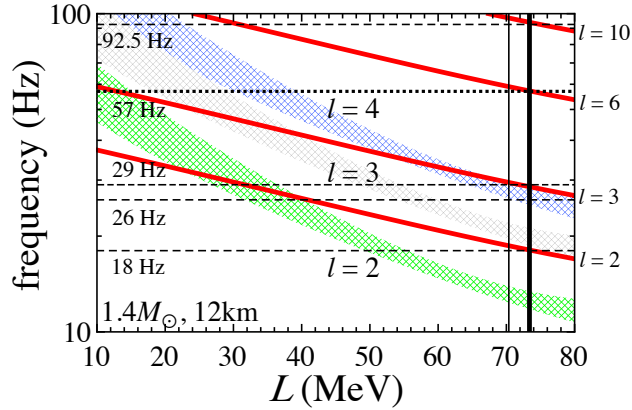}
\end{center}
\caption{
(Color online)
The $\ell=2$, 3, and 4 fundamental frequencies (painted regions)
of the torsional oscillations in the layer of the tube and bubble 
phases, calculated as a function of $L$ for the neutron star 
models with $M=1.4M_\odot$ and $R=12$ km.  The lower and upper 
boundaries of the painted regions correspond to the results obtained
for the maximum enthalpy (${\cal R}=1$) and the minimum enthalpy
(${\cal R}=0$), respectively.  For reference, the low-lying QPO 
frequencies derived by the conventional non-Bayesian analysis for
SGR 1806-20 are shown by horizontal lines.   The QPO frequencies 
except 26 Hz can be interpreted as manifestations of the 
$\ell=2$, 3, 6, and 10 fundamental torsional oscillations that are
excited in the layer composed of spherical and cylindrical nuclei 
\citep{SIO2018}, as illustrated by the solid lines that 
denote the corresponding eigenfrequencies obtained by assuming that 
the dripped neutron outside the cylindrical nuclei do not 
participate in the oscillations (minimum enthalpy, i.e., 
$N_s/N_d=1$). The vertical thick line, i.e., $L=73.4$ MeV, denotes 
the optimal value of $L$ for explaining the observed QPOs except 
the 26 Hz QPO in terms of the torsional oscillations in the 
sphere-cylinder layer with minimum enthalpy, while the vertical thin line, 
i.e., $L=70.4$ MeV, denotes the corresponding value of $L$ 
in the case of maximum enthalpy, i.e., $N_s/N_d=0$ \citep{SIO2018}.
}
\label{fig:M14R12}
\end{figure}

Let us now illustrate how the newly examined torsional
oscillations in the tube-bubble layer could be accommodated
to the QPO observations of frequencies lower than 100 Hz,
including 26 Hz, for typical $M$-$R$ sets of the stellar models.   
For $M=1.4M_\odot$ and $R=12$ km, such illustration can be seen 
from Fig.~\ref{fig:M14R12}, in which the 18, 29, 57, and 
92.5 Hz QPOs in SGR 1806-20 are as usual identified as
the $\ell=2$, 3, 6, and 10 fundamental frequencies in the 
sphere-cylinder layer, whereas the 26 Hz QPO, which is difficult 
to explain in terms of the oscillation in the sphere-cylinder layer, 
can reasonably be identified as the $\ell=4$ fundamental
frequency in the tube-bubble layer.  We remark that the
optimal value of $L$ for explaining the observed 
low-lying QPOs ranges between 70.4 and 73.4 MeV
for neutron stars with $M=1.4M_\odot$ and $R=12$ km.

\begin{figure*}
\begin{center}
\begin{tabular}{cc}
\includegraphics[scale=0.5]{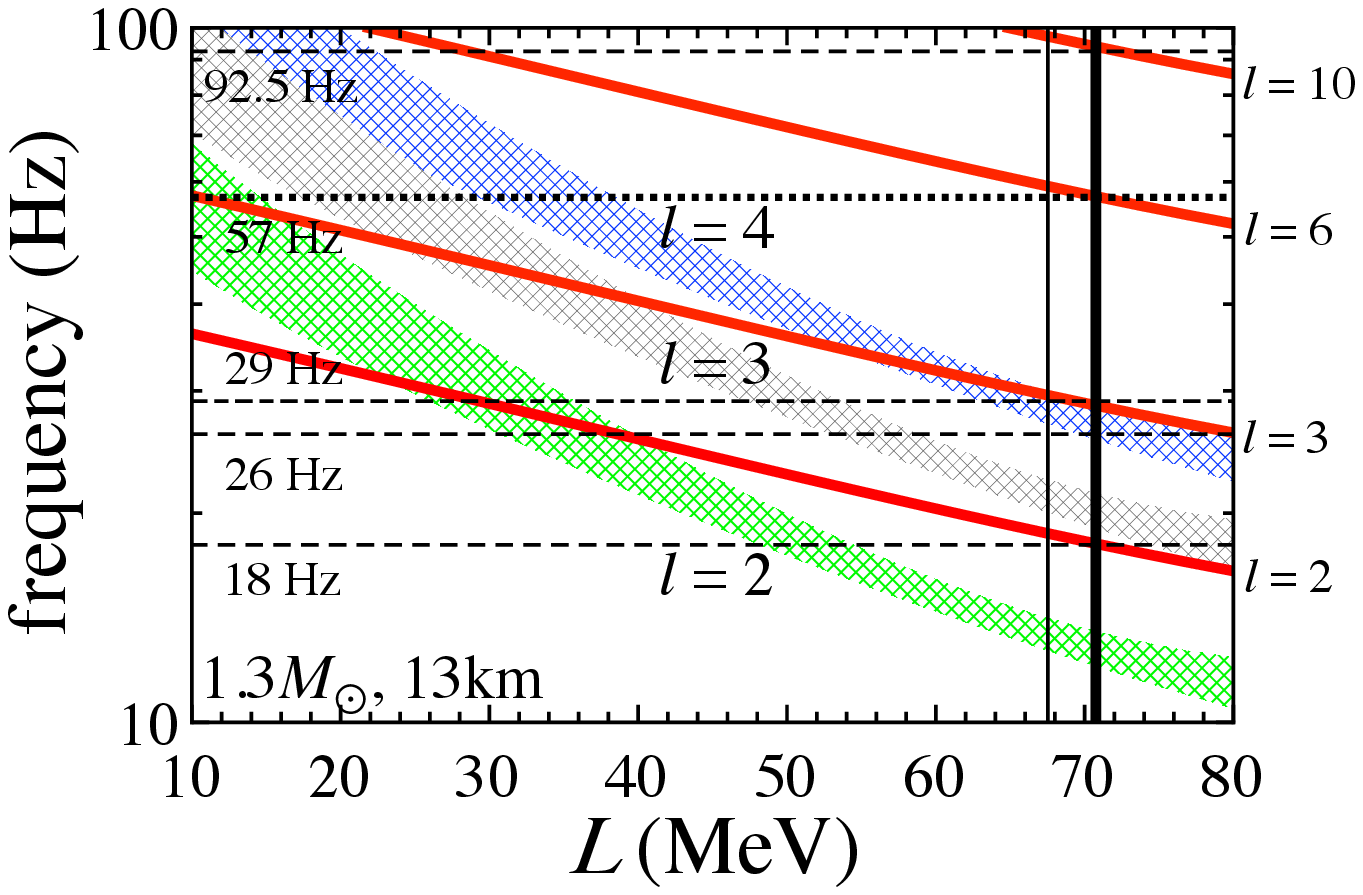} &
\includegraphics[scale=0.5]{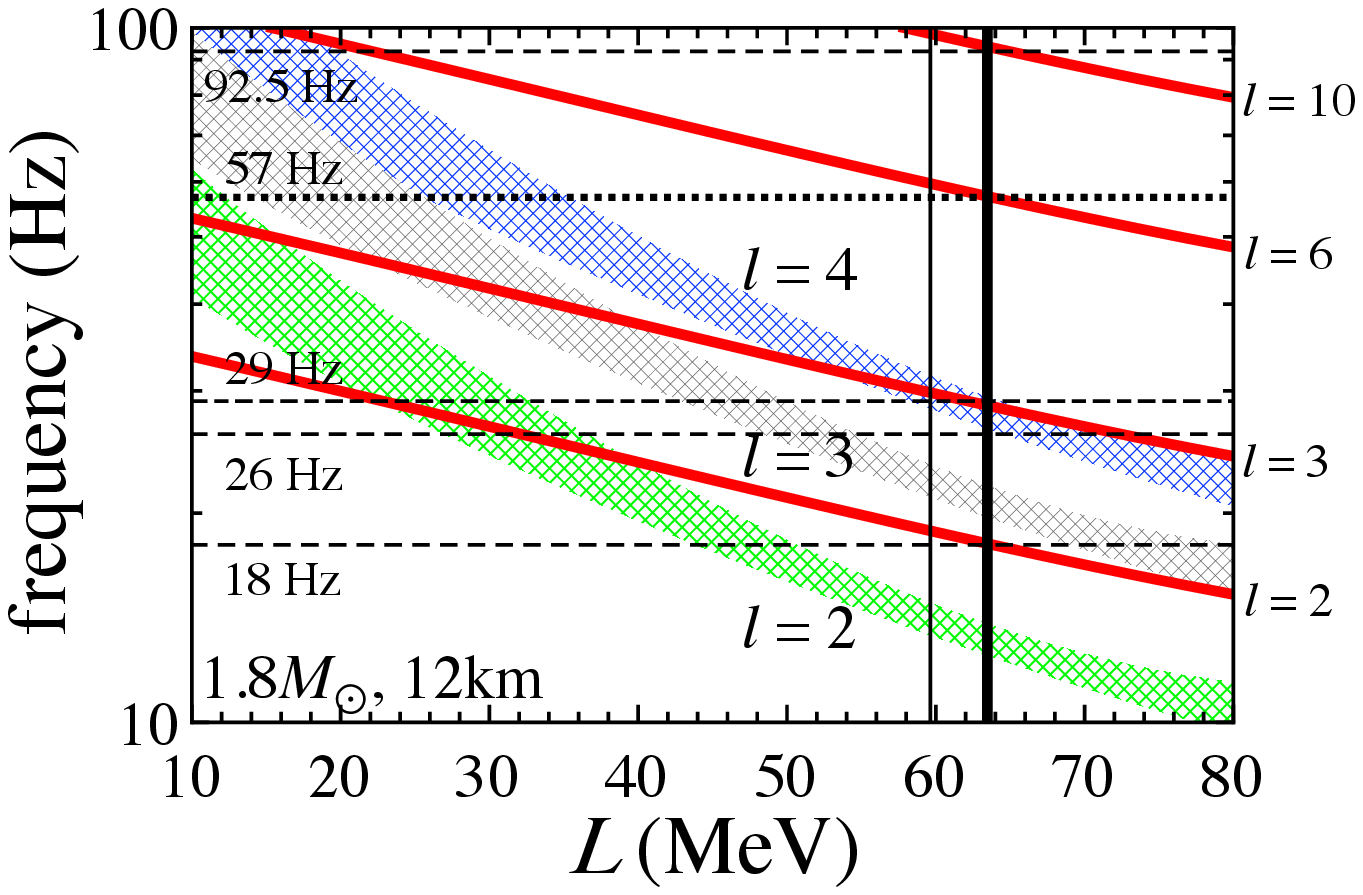}  
\end{tabular}
\end{center}
\caption{
(Color online)
Same as Fig.~\ref{fig:M14R12}, but for the neutron star models 
with $M=1.3M_\odot$ and $R=13$ km in the left panel and 
with $M=1.8M_\odot$ and $R=12$ km in the right panel. 
The optimal values of $L$ denoted by the vertical 
thick and thin lines are $L=70.8$ and 67.5 MeV, 
respectively, in the left panel and $L=63.5$ and 59.6 MeV,
respectively, in the right panel.
}
\label{fig:M13M18}
\end{figure*}

We then examine whether or not the above-mentioned identification,
which strictly holds for $(M, R)=(1.4M_\odot,12 {\rm km})$, still works
out for other sets of $(M, R)$.  For neutron star models with 
$(M, R)=(1.3M_\odot,13 {\rm km})$ and $(1.8M_\odot,12{\rm km})$, 
we again calculate the eigenfrequencies of the double-layer torsional
oscillations, as shown in Fig.~\ref{fig:M13M18}.  We find that 
the 18, 29, 57, and 92.5 Hz QPOs in SGR 1806-20 can be still consistent 
with the $\ell=2,3,6$, and 10 fundamental frequencies in the sphere-cylinder
layer for such a range of the optimal $L$ as 67.5--70.8 MeV and 
59.6--63.5 MeV, respectively.  This shift of the optimal $L$ could
open up an opportunity of selecting $M$ and/or $R$ because 
the $L$ dependence of the fundamental frequencies 
in the sphere-cylinder layer is different from that 
in the tube-bubble layer.   
In fact, one can observe the tendency that the more 
massive neutron star, the more difficult to explain
the 26 Hz QPO in terms of the $\ell=4$ fundamental 
oscillation in the tube-bubble layer, 
as long as we adopt the optimal value of $L$ that enables us to identify
the 18, 29, 57, and 92.5 Hz QPOs as the oscillations in the 
sphere-cylinder layer. Note that the 
fundamental frequencies scale as $R^{-1}$ both in the tube-bubble 
and sphere-cylinder layers, implying that $R$ is not constrained 
in the present approach.  We can thus conclude that 
light neutron star models are favored over 
heavy ones in our identification.  Incidentally, the 
$(M, R)=(1.3M_\odot,13 {\rm km})$ case is consistent with 
the neutron star models considered to be relevant
as a result of the comparison of the constraint on $K_0$, 
which is obtained by assuming that the lowest 
overtone frequency in the sphere-cylinder layer is 
equal to the QPO frequency of 626.5 Hz observed from 
SGR 1806-20, with the terrestrial constraint on $K_0$ 
\citep{SIO2018}.  Furthermore, the 
$(M, R)=(1.3M_\odot,13 {\rm km})$ case is consistent with 
the mass and radius formulas for low-mass neutron 
stars \citep{SIOO2014}, given the optimal 
value of $L\sim 70$ MeV, and also with the constraint 
on the mass and radius of each of the merging binary
neutron stars \citep{A2018}.

\begin{figure}
\begin{center}
\includegraphics[scale=0.6]{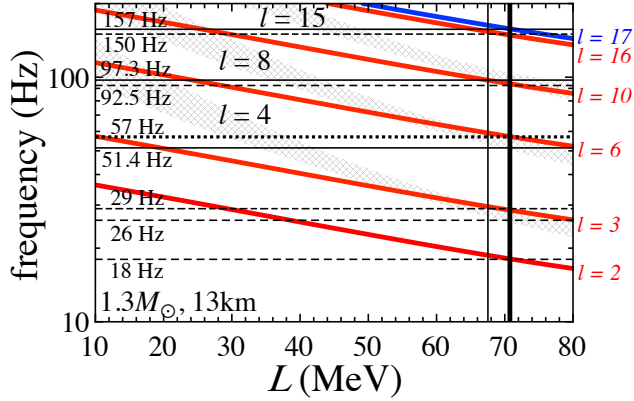}
\end{center}
\caption{
(Color online) Relations between the newly found 
QPOs of 51.4, 97.3, and 157 Hz in SGR 1806-20
\citep{MCS18}, which are shown by horizontal solid lines, 
and a selected set of the crustal torsional oscillations
for the neutron star models with $M=1.3M_\odot$ and $R=13$
km.  The 51.4 and 97.3 Hz QPOs are identifiable as the 
$\ell=8$ and 15 fundamental torsional oscillations in the 
tube-bubble layer, while the 157 Hz QPO is identifiable as 
the $\ell=17$ fundamental torsional oscillations in the 
sphere-cylinder layer.  The dashed and dotted lines 
denote the originally discovered QPOs, which except for the 26 Hz QPO
have already been identified by us as manifestations of the fundamental
torsional oscillations in the sphere-cylinder layer, while the 26 Hz QPO
is identified as the $\ell=4$ oscillation in the tube-bubble layer as mentioned in text.
}
\label{fig:M13R13c}
\end{figure}

Thanks to the smaller shear modulus in the tube phase, 
which leads to the smaller fundamental frequencies in the 
tube-bubble layer than those in the sphere-cylinder
layer, we have a chance to explain not only the originally
discovered QPOs but also the QPOs newly found by 
a Bayesian procedure, e.g., the 51.4, 97.3, and 157 Hz 
QPOs in SGR 1806-20 \citep{MCS18}\footnote{In appendix 
\ref{sec:appendix_1}, we tabulate a possible 
correspondence between the crustal torsional oscillations and 
all the 26 QPOs shown in Table 1 in \cite{MCS18}.}.  
In practice, we illustrate the identification of 
these three QPOs for the neutron star models with 
$M=1.3M_\odot$ and $R=13$ km in Fig.~\ref{fig:M13R13c}.
As already shown in \cite{SIO2018}, the frequencies of 18, 29, 
57, 92.5, and 150 Hz can be identified as the 
$\ell=2$, 3, 6, 10, and 16 fundamental frequencies in
the sphere-cylinder layer.  In a similar way, we find 
that the newly found QPO of 157 Hz can also be
identified as the $\ell=17$ fundamental frequency, while 
the newly found QPOs of 51.4 and 97.3 Hz
can be identified as the fundamental oscillations in 
the tube-bubble layer, as is the case with 
the 26 Hz QPO, in such a way that 26, 51.4, and 97.3 Hz 
correspond to the $\ell=4$, 8, and 15 fundamental oscillations 
in the tube-bubble layer.

\section{Conclusion}
\label{sec:V}

We have calculated the eigenfrequencies of the torsional oscillations in
the tube-bubble layer, in contrast to our previous work in which we 
calculated those only in the bubble layer, and successfully identified
the newly found QPOS as the fundamental oscillations either in 
the tube-bubble or sphere-cylinder layer.  In the course of the 
calculations, we find that the shear modulus, which characterizes the 
torsional oscillations, decreases in the tube phase as the slope
parameter $L$ increases.  As a result, the fundamental frequencies in the 
tube-bubble layer can become smaller than those in the 
sphere-cylinder layer.  We also find that the fundamental frequencies 
in the tube-bubble layer can be parameterized as a function of $L$,
and that the dependence on $L$ is different from that obtained for 
the fundamental frequencies in the sphere-cylinder layer. 
Remarkably, such a different dependence on $L$ helps us to explain 
not only the QPO frequencies originally discovered in SGR 1806-20 but 
also those newly found in the same object by a Bayesian procedure 
in terms of the eigenfrequencies of the fundamental torsional oscillations 
either in the tube-bubble or sphere-cylinder layer of a relatively low
mass neutron star constructed from the EOS of $L\sim 70$ MeV. 
We also remark that such a neutron star model and the suitable value 
of $L$ are consistent with the mass and radius formulas of low-mass 
neutron stars and the constraint from the gravitational waves from the 
neutron star binary merger, GW170817.

As a possible extension of this study, it would be of interest to 
analyze the QPO widths, which could give us information of the internal
magnetic structure via possible coupling of the crustal torsional
oscillations with the Alfv\'{e}n continuum in the core \citep{MCS18}.  
Generally, magnetars are considered to have a toroidal field that is by 
an order of magnitude higher than the poloidal field.  The question of 
whether or not this picture is relevant might be possibly answered.

This work was supported in part by Grants-in-Aid for Scientific Research 
through Grant Nos.\ 17K05458, 18H01211, and 18H05406
provided by Japan Society for the Promotion of Science (JSPS).

\appendix
\section{Possible identifications of newly found QPOs as 
the crustal torsional oscillations}   
\label{sec:appendix_1}

\begin{figure*}
\begin{center}
\begin{tabular}{cc}
\includegraphics[scale=0.55]{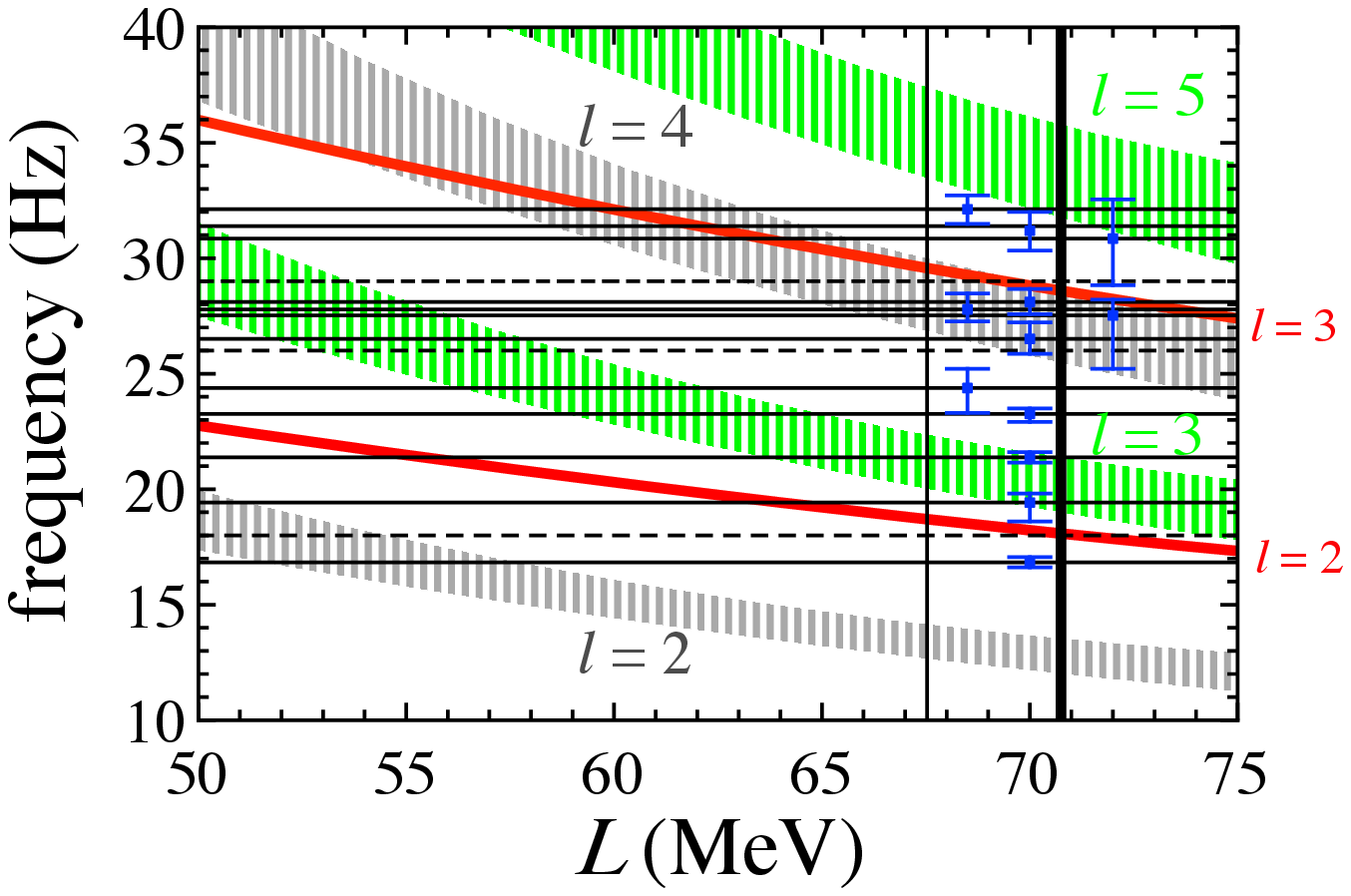} &
\includegraphics[scale=0.55]{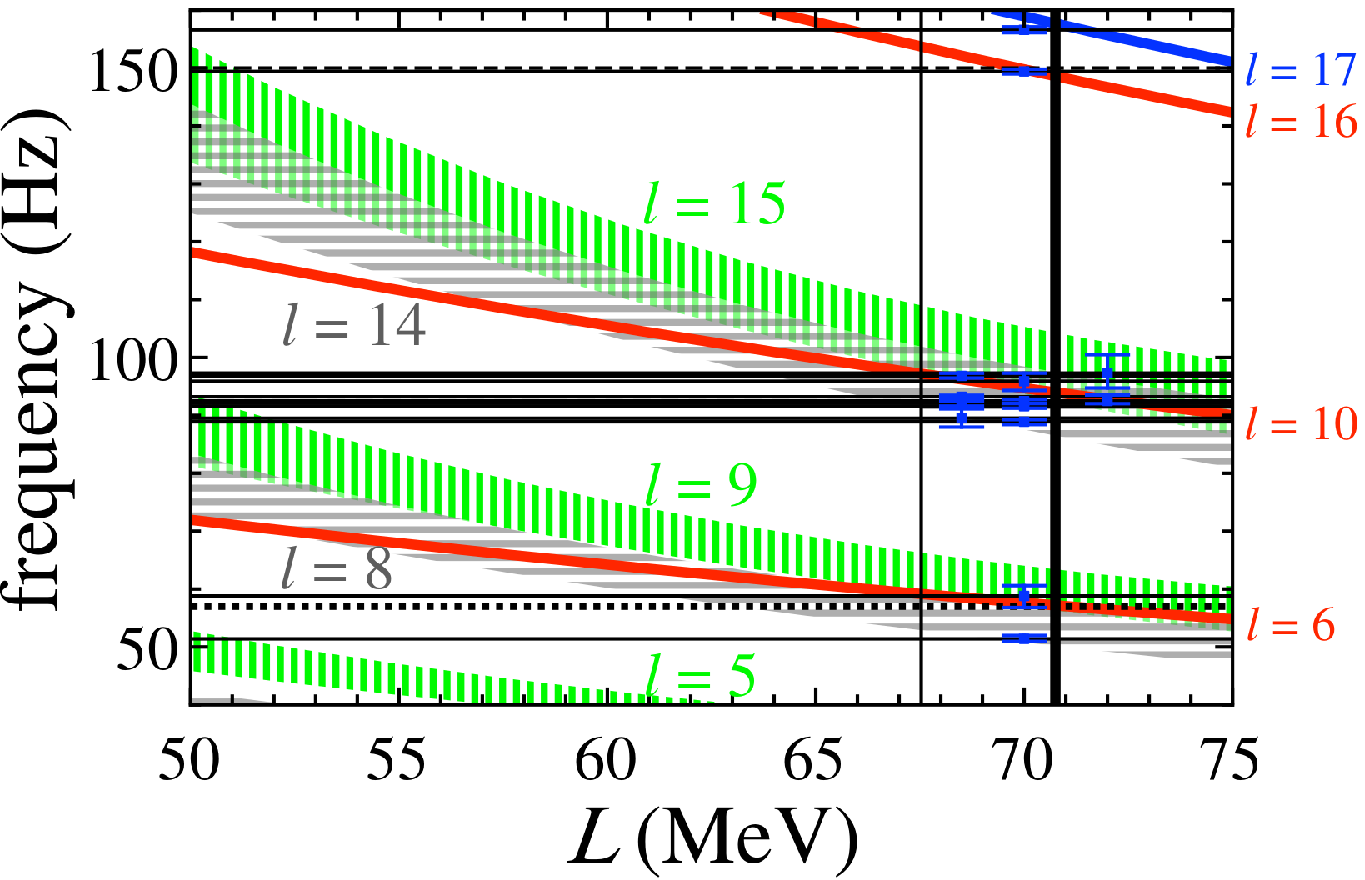}  
\end{tabular}
\end{center}
\caption{
(Color online)
Comparison between the fundamental frequencies of the crustal torsional 
oscillations and the QPO frequencies found in SGR 1806-20 
in the low frequency region (left panel) and in the high 
frequency region (right panel).  The dashed and dotted lines denote the 
QPO frequencies discovered in earlier studies, while the 
solid lines denote the QPO frequencies newly found by a 
Bayesian analysis \citep{MCS18}.  With respect to the newly found 
QPOs, we also display the $1\sigma$ uncertainties at the
position of either $L=68.5$, 70, or 72 MeV.  For the neutron star model 
with $M=1.3M_\odot$ and $R=13$ km, the fundamental frequencies in 
the sphere-cylinder layer are denoted by solid lines, while those 
in the tube-bubble layer are by painted regions. 
The possible correspondence between the observed QPO frequencies
and the fundamental frequencies is shown in Table \ref{tab:QPO}. 
}
\label{fig:M13M13}
\end{figure*}

In this Appendix, we attempt overall identifications of the 26 QPOs in 
SGR 1806-20 newly found by a Bayesian analysis that adopts the 
signals with Bayes factors larger than 1000 and frequency widths less 
than 10 Hz as QPOs \citep{MCS18}, together with the six QPOs 
originally discovered in SGR 1806-20, as the crustal torsional 
oscillations.  Although the originally discovered QPOs are 
expected to be more robust than those newly found by the Bayesian 
procedure, it would be convincing if most of the newly found 26 
QPOs could also be identified as the fundamental torsional 
oscillations with various values of $\ell$  that are excited either
in the sphere-cylinder layer or in the tube-bubble layer. 
The comparison between the QPO frequencies and the fundamental 
frequencies of the crustal torsional oscillations is shown in Fig.\ 
\ref{fig:M13M13}.  Most of the newly found QPOs, except for a few 
lowest-lying ones, turn out to be reasonably identifiable, as shown in 
Table \ref{tab:QPO}.

In Table \ref{tab:QPO}, we also list the frequency widths, $\Delta f$, 
given by \cite{SW2006,QPO2,MCS18} in the second column, where the 
values of $\Delta f$ more than 1.5 Hz among the newly found QPOs 
are shown in boldface.  One can observe from this table that 
all the QPOs with relatively large frequency widths
are identifiable as the torsional oscillations in the tube-bubble
layer.  This might suggest that the torsional oscillations in 
such a deep region of the crust are easy to undergo Landau 
damping by coupling with the neighboring magnetic oscillations 
in the core.

In addition, some of the newly found QPOs, which are rather 
difficult to identify, are marked with the question mark in Table 
\ref{tab:QPO}.  These are all low-lying and could be marginally 
identifiable as shown in the table.  As an exception, the 23.26 Hz QPO seems 
to be unidentifiable within our framework.  Moreover, we should remark 
that the possible presence of the 9.2 Hz QPO in SGR 1806-20 was 
deduced from another Bayesian analysis \citep{PGSE2018}, which is 
in fact difficult to understand within our framework.  For
future research involved, it is interesting to note that these two QPOs, 
of which the origin needs to be clarified if any, have a frequency 
(23.26 Hz) that is almost twice as high as the $\ell=2$ fundamental 
frequency in the tube-bubble layer and a frequency (9.2 Hz) that
is almost half as high as the $\ell=2$ fundamental frequency 
in the sphere-cylinder layer.

\begin{table}
\centering
\begin{minipage}{65mm}
\caption{
Possible correspondence of the QPOs found in SGR 1806-20 to the fundamental 
torsional oscillations with various values of $\ell$ either in the 
sphere-cylinder (SP+C) layer or in the tube-bubble (CH+SH) layer. 
The frequency width of the respective QPO, derived from the X-ray data
\citep{SW2006,QPO2,MCS18}, is also shown in the second column.  The top 6 
QPOs were reported in earlier publications, while the other 26 QPOs 
were newly found.  Among the newly found QPOs, the frequency 
widths larger than 1.5 Hz are shown in boldface.
}
\begin{tabular}{cccc}
\hline\hline
  QPO (Hz) & $\Delta f$ (Hz) & SP+C & CH+SH   \\
\hline
  18    & $1.9\pm 0.2$ &  $\ell=2$  &                \\
  26    & $3.0\pm 0.2$ &                 & $\ell=4$  \\
  29    & $4.1\pm 0.5$ &  $\ell=3$  &                \\
  57    & 4.4                 &  $\ell=6$  &                \\
  92.5 & $1.7^{+0.7}_{-0.4}$ & $\ell=10$ &      \\
  150 & $17\pm 5$ & $\ell=16$  &                    \\
\hline
  16.83 & 0.51 &  $\ell=2$ (?) &                       \\
  19.42 & {\bf 3.42} &             & $\ell=3$          \\
  21.38 & 0.51 &                     & $\ell=3$ (?)    \\
  23.26 & 0.51 &          ?         &  ?                    \\
  24.38 & {\bf 4.19} &             & $\ell=4$ (?)     \\
  26.51 & {\bf 2.03} &             & $\ell=4$           \\
  27.52 & 1.28 & $\ell=3$       &                        \\
  27.78 & 1.24 & $\ell=3$       &                        \\
  28.11 & 1.38 & $\ell=3$      &                         \\
  30.85 & {\bf 5.43} &             &  $\ell=5$          \\
  31.19 & {\bf 1.90} &             &  $\ell=5$          \\
  32.12 & {\bf 1.93} &            &  $\ell=5$           \\
  51.40 & 0.52 &                    & $\ell=8$             \\
  58.81 & {\bf 5.36} & ($\ell=6$)      &  $\ell=9$   \\
  88.90 & 0.73 &                   & $\ell=14$           \\
  89.51 & {\bf 4.78} &           & $\ell=14$            \\
  91.58 & 1.14 & $\ell=10$  &                          \\
  92.09 & 0.82 & $\ell=10$  &                          \\
  92.47 & 0.69 & $\ell=10$  &                          \\
  92.49 & 0.71 & $\ell=10$  &                          \\
  93.21 & 0.50 & $\ell=10$  &                          \\
  95.90 & {\bf 3.74} &           & $\ell=15$  \\
  96.82 & 0.60 &                   & $\ell=15$  \\
  97.31 & {\bf 5.94} &           & $\ell=15$  \\
  149.41 & 0.79 & $\ell=16$  &   \\
  156.59 & 1.14 & $\ell=17$  &   \\
\hline\hline
\end{tabular}
\label{tab:QPO}
\end{minipage}
\end{table}


\end{document}